# Superconductivity in trilayer nickelate La$_4$Ni$_3$O$_{10}$ under pressure

Mingxin Zhang[1#], Cuiying Pei[1#], Di Peng[2,3#], Xian Du[4#], Weixiong Hu[1#], Yantao Cao[5,6#], Qi Wang[1,7], Juefei Wu[1], Yidian Li[4], Huanyu Liu[1], Chenhaoping Wen[1], Jing Song[8], Yi Zhao[1], Changhua Li[1], Weizheng Cao[1], Shihao Zhu[1], Qing Zhang[1,9], Na Yu[1], Peihong Cheng[1], Lili Zhang[10], Zhiwei Li[5], Jinkui Zhao[6,] Yulin Chen[1,7,11], Changqing Jin[8,12], Hanjie Guo[6*], Congjun Wu[13,14], Fan Yang[15*], Qiaoshi Zeng[2,3*], Shichao Yan[1,7*], Lexian Yang[4*], Yanpeng Qi[1,7,9*]

1. School of Physical Science and Technology, ShanghaiTech University, Shanghai 201210, China
2. Shanghai Key Laboratory of Material Frontiers Research in Extreme Environments (MFree), Institute for Shanghai Advanced Research in Physical Sciences (SHARPS), Shanghai 201203, China
3. Center for High Pressure Science and Technology Advanced Research, Shanghai 201203, China
4. State Key Laboratory of Low Dimensional Quantum Physics, Department of Physics, Tsinghua University, Beijing 100084, China
5. Key Lab for Magnetism and Magnetic Materials of the Ministry of Education, School of Physical Science and Technology, Lanzhou University, Lanzhou 730000, Gansu, China
6. Songshan Lake Materials Laboratory, Dongguan 523808, Guangdong, China
7. ShanghaiTech Laboratory for Topological Physics, ShanghaiTech University, Shanghai 201210, China
8. Beijing National Laboratory for Condensed Matter Physics, Institute of Physics, Chinese Academy of Sciences, 100190, Beijing, China
9. Shanghai Key Laboratory of High-resolution Electron Microscopy, ShanghaiTech University, Shanghai 201210, China
10. Shanghai Synchrotron Radiation Facility, Shanghai Advanced Research Institute, Chinese Academy of Sciences, Shanghai 201203, China
11. Department of Physics, Clarendon Laboratory, University of Oxford, Parks Road, Oxford OX1 3PU, UK
12. School of Physical Sciences, University of Chinese Academy of Sciences, 100190, Beijing, China
13. Institute for Theoretical Sciences, Westlake University, Hangzhou 310024, Zhejiang, China
14. New Cornerstone Science Laboratory, Department of Physics, School of Science, Westlake University, Hangzhou 310024, Zhejiang, China

15. School of Physics, Beijing Institute of Technology, Beijing 100081, China

\# These authors contribute equally to this work.

\* Correspondence should be addressed to Y.P.Q. (qiyp@shanghaitech.edu.cn) or L.X.Y(lxyang@tsinghua.edu.cn) or S.C.Y(yanshch@shanghaitech.edu.cn) Q.S.Z. (zengqs@hpstar.ac.cn) or F.Y (yangfan_blg@bit.edu.cn) or H.J.G (hjguo@sslab.org.cn)

**ABSTRACT**

**Nickelate superconductors have attracted a great deal of attention over the past few decades due to their similar crystal and electronic structures with high-temperature cuprate superconductors. Here, we report the superconductivity in a pressurized Ruddlesden-Popper phase single crystal, $La_4Ni_3O_{10}$ ($n$ = 3), and its interplay with the density wave order in the phase diagram. With increasing pressure, the density wave order as indicated by the anomaly in the resistivity is progressively suppressed, followed by the emergence of the superconductivity around 25 K under *I4/mmm* space group. The susceptibility measurements confirm bulk superconductivity with a volume fraction exceeding 80%. Moreover, theoretical analysis unveils that antiferromagnetic (AFM) super-exchange interactions can serve as the effective pairing interaction for the emergence of superconductivity (SC) in pressurized $La_4Ni_3O_{10}$. Our research provides a new platform for the investigation of the unconventional superconductivity mechanism in Ruddlesden–Popper trilayer perovskite nickelates.**

# I.INTRODUCTION

The discovery of superconductivity in cuprates has opened the avenue to explore high-temperature (high-$T_c$) superconductivity[1, 2]. Almost all the discovered cuprates superconductors so far share a common two-dimensional $CuO_2$ plane and half-filled Cu $3d^9$ electronic configuration, which are believed to be essential to the emergence of high-$T_c$ superconductivity[3, 4]. Besides cuprates, iron-based superconductors are another prime material family with superconductivity well above the McMillan limit at ambient pressure[5, 6]. In contrast to the cuprates, iron-based superconductors with Fe $3d^6$ electronic configuration have a multi-band and multi-orbital nature[7, 8]. Despite sustained efforts that have been devoted more than three decades, the full understanding of their superconductivity mechanism remains a challenging and long-sought problem in condensed matter physics.

Nickelates are considered to be an ideal candidate for exploring new cuprate-like high-$T_c$ superconductivity due to the proximity of Ni$^+$(3$d^9$) electron configuration[9]. However, superconductivity was not experimentally realized in nickelates until 2019, when *Li* et al. discovered superconductivity within the thin films of infinite-layer nickelates Nd$_{1-x}$Sr$_x$NiO$_2$ with $T_c$ ~ 9-15 K[10, 11], which quickly promoted renewed interest in the exploration of nickel-based superconductors. Subsequently, $T_c$ has been extended to other infinite-layer nickelates with different rare-earth elements [12-15]. However, superconductivity is limited to thin film materials. Structurally, whether there exists a nickelate superconductor family with bulk superconductivity is still an open question.

Recently, the experimental breakthrough revealed high-$T_c$ superconductivity above liquid-nitrogen temperature in pressurized bulk La$_3$Ni$_2$O$_7$ single crystals, which is one of the Ruddlesden-Popper (RP) phases La$_{n+1}$Ni$_n$O$_{3n+1}$ ($n$ = 2) with the valence of Ni$^{2.5+}$ [16]. The synchrotron X-ray diffraction (XRD) patterns and Density functional theory (DFT) calculations indicate that the La$_3$Ni$_2$O$_7$ undergoes a structural phase transition from *Amam* to *I4/mmm* with the emergence of superconductivity, and a series of follow-up experiments have elaborated that oxygen defects and hydrostatic environment are essential for the discovery of superconductivity[17-23]. Moreover, many theoretical works are put forward to understand the underlying mechanisms of high-$T_c$ superconductivity in La$_3$Ni$_2$O$_7$[24-39].

Among RP phase nickelates, the trilayer La$_4$Ni$_3$O$_{10}$ ($n$ = 3) has drawn more attention due to the metallic ground state and metal-to-metal transition at intermediate temperatures[40-43]. The nominal valence state of Ni is +2.66 in $La_4Ni_3O_{10}$, which is usually believed to be given by mixed valence states of $Ni^{2+}$ ($3d^8$) and $Ni^{3+}$ ($3d^9$). Considering the similar crystal structure and Ni valence with $La_3Ni_2O_7$, it is natural to consider investigating the superconductivity and electronic structure in pressurized $La_4Ni_3O_{10}$ single crystals.

In this work, we grow high-quality $La_4Ni_3O_{10}$ single crystal with an optical-image floating zone furnace under high oxygen pressure and perform high-pressure transport measurements using NaCl as the pressure-transmitting medium to improve the hydrostatic environment. We find superconductivity with a $T_c$ of 25 K in $La_4Ni_3O_{10}$ single crystals under pressure. A distinctive diamagnetic transition consists with the resistance measurements, further confirming the emergence of superconductivity in $La_4Ni_3O_{10}$ single crystals. By combining scanning tunneling microscopy/spectroscopy, angle-resolved photoemission spectroscopy, and *ab initio* calculation, we systematically investigate the electronic structure of trilayer nickelate $La_4Ni_3O_{10}$. These observations broaden and contribute to the superconducting family of nickelates.

## II. RESULTS AND DISCUSSION

### A. Trilayer nickelate.

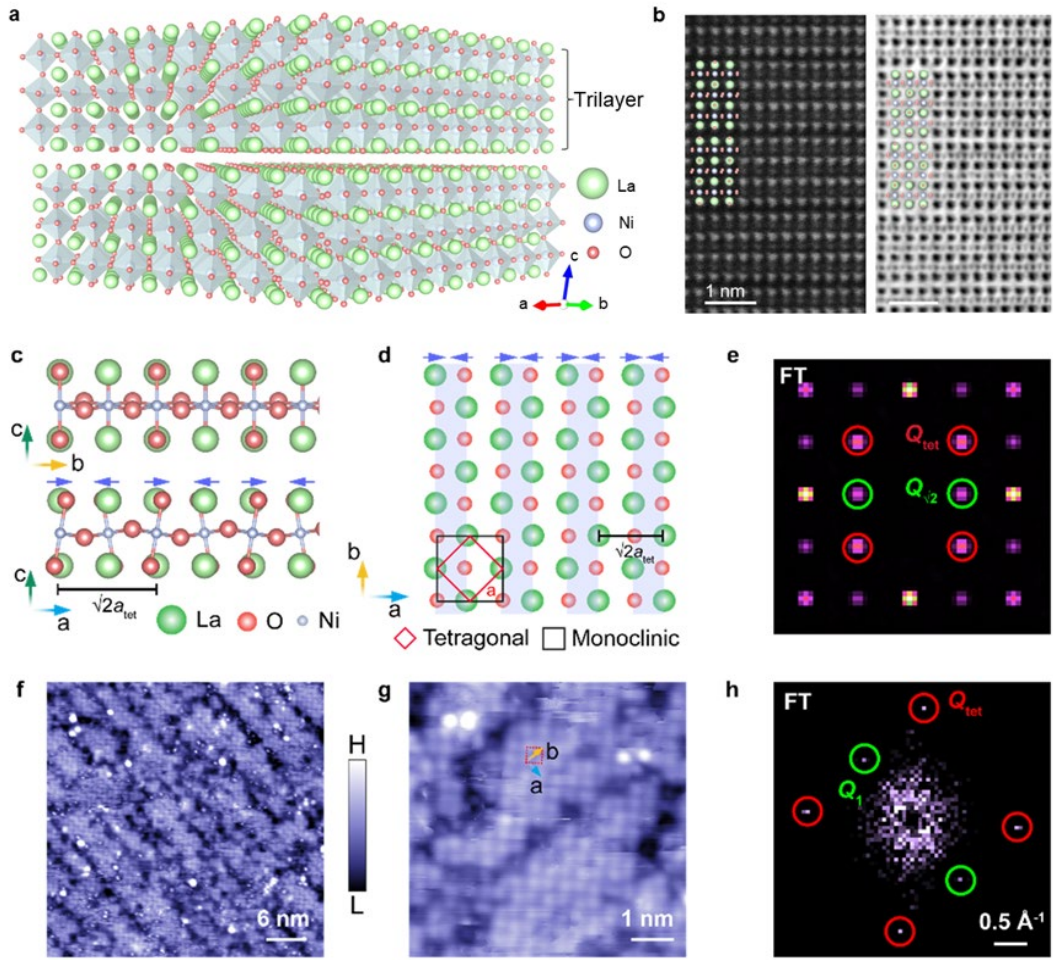

**FIG .1. Crystal structure and surface topographies of $La_4Ni_3O_{10}$ single crystals.** **a**, schematic illustration of the crystal structure of $La_4Ni_3O_{10}$ crystal. **b**, high-angle annular dark-field (left) and annular bright-field (right) STEM images of $La_4Ni_3O_{10}$ along the b axis. **c**, side-view crystal structures along the *ac* plane (upper) and the *bc* plane (lower), respectively. **d**, the top-view crystal structure of the LaO surface. The appended red and black squares are the unit cells for the tetragonal and monoclinic phases. The light blue ribbons indicate the unidirectional reconstruction. **e**, FT image of the schematic crystal structure shown in d. The red and green circles indicate the tetragonal Bragg lattice ($Q_{tet}$) and the unidirectional structural distortion pattern ($Q_{\sqrt{2}}$), respectively. **f, g**, Constant-current STM topographies taken on the LaO surface. The colored arrows in g indicate the monoclinic lattice. **h**, FT image of the STM topography shown in g.

La$_4$Ni$_3$O$_{10}$, one of the RP phases La$_{n+1}$Ni$_n$O$_{3n+1}$ ($n$ = 3) nickelates, crystallizes in a monoclinic structure ($a$ = 5.41 Å, $b$ = 5.47 Å, $c$ = 14.22 Å), consisting of 3 layers of perovskite-type LaNiO$_3$, separated by a single rocksalt-type LaO layer along the crystallographic $c$-axis direction (Fig. 1a). As shown in Fig. 1b, high-angle annular dark-field (left) and annular bright-field (right) transmission electron microscopy images (Fig. 1b) illustrates each unit contains well-ordered three layers of inclined NiO$_6$ octahedral layers stacking alone the $c$-axis direction. The inclination of the NiO$_6$ octahedra leads to the bucking of the Ni-O planes. The side-view crystal structure of the La-Ni-O layer in the monoclinic phase of La$_4$Ni$_3$O$_{10}$ is shown in Fig. 1c. As we can see, the Ni-O bonds are tilted in the $bc$ plane instead of the $ac$ plane, resulting in a unidirectional structural reconstruction compared to that of the tetragonal phase. Figure 1d shows the in-plane unit cells for the tetragonal and the monoclinic phases. The size of the reconstructed monoclinic unit cell is twice that of the tetragonal unit cell due to the structural reconstruction. The unidirectional reconstruction results in a stripe-like structure as illustrated by the light blue ribbons in Fig. 1d. The periodicity of the stripe is √2 times of the tetragonal lattice $a_{tet}$, as indicated by the $Q_{\sqrt{2}}$ in the Fourier transform (FT) of the NiO lattice (Fig. 1e).

The unidirectional reconstruction in La$_4$Ni$_3$O$_{10}$ can be confirmed by the low-temperature scanning tunneling microscopy (STM) measurements. Figure 1f shows a typical STM topography taken on the large-area LaO-terminated surface. On the LaO surface, we observe an ordered square lattice with some adatom-like impurities and several rip-like dark stripes along the diagonal direction of the square lattice. These features can be seen more clearly in the zoom-in STM topography as shown in Fig. 1g. The square lattice is measured to be 0.40 nm, which is consistent with the in-plane LaO tetragonal lattice as shown in Fig. 1d. This indicates that the rip-like dark stripes are along the reconstruction direction. To further quantify the reconstructed surface, we perform FT to the STM topography (Fig. 1h). In the FT image, apart from the $Q_{tet}$ originated from LaO in-plane tetragonal lattice, there is a set of unidirectional $Q_1$ vector, which is $1/\sqrt{2}$ times of the $Q_{tet}$. The length and direction of the $Q_1$ wave vector are the same as the $Q_{\sqrt{2}}$ in Fig. 1e, indicating that there exists a stripe-like unidirectional structural reconstruction on the LaO surface. Interestingly, the rip-like dark stripes in the real space are also along the reconstructed stripe direction, and they can contribute broader signals in the FT images (Fig. 1g, also see Fig. S3). This suggests that the rip-like dark stripes might be related to the reconstruction in $La_4Ni_3O_{10}$ due to the combination of the unidirectional strain-like effect and the local oxygen vacancies[22]. Although the unidirectional reconstruction has been revealed, we find no clear signs of the charge density wave pattern in the STM topographies on the LaO-terminated surface.

### B. Transport under ambient pressure.

Figures 2a-c show the resistivity, magnetization, and heat capacity measured on $La_4Ni_3O_{10}$ single crystals. The resistivity decreases with cooling temperature from 300 K, revealing typical metallic behavior. A pronounced anomalous peak at around 130 K, suggests possible density-wave states. The in-plane magnetization was measured under a magnetic field of 0.4 T[41, 44], which shows a sharp decrease at the $T^*$~132 K, similar to typical charge or spin density wave materials such as $K_{0.3}MoO_3$[45], $CsV_3Sb_5$[46] and $LaFeAsO$[5]. The temperature-dependent heat capacity also exhibits an anomalous peak at the same temperature, and it can be well fitted by the formula $C_p/T = \gamma + \beta T^2$, where $\gamma$ is the Sommerfeld coefficient. The fit yields $\gamma = 13.5$ mJ mole$^{-1}$K$^{-2}$ and $\beta = 0.34$ mJ mole$^{-1}$ K$^{-4}$, which agrees with the previous measurement [41]. The Debye temperatures and density of states $N(E_F)$ can be estimated as 459 K and 2.1 states eV$^{-1}$ per Ni, respectively.

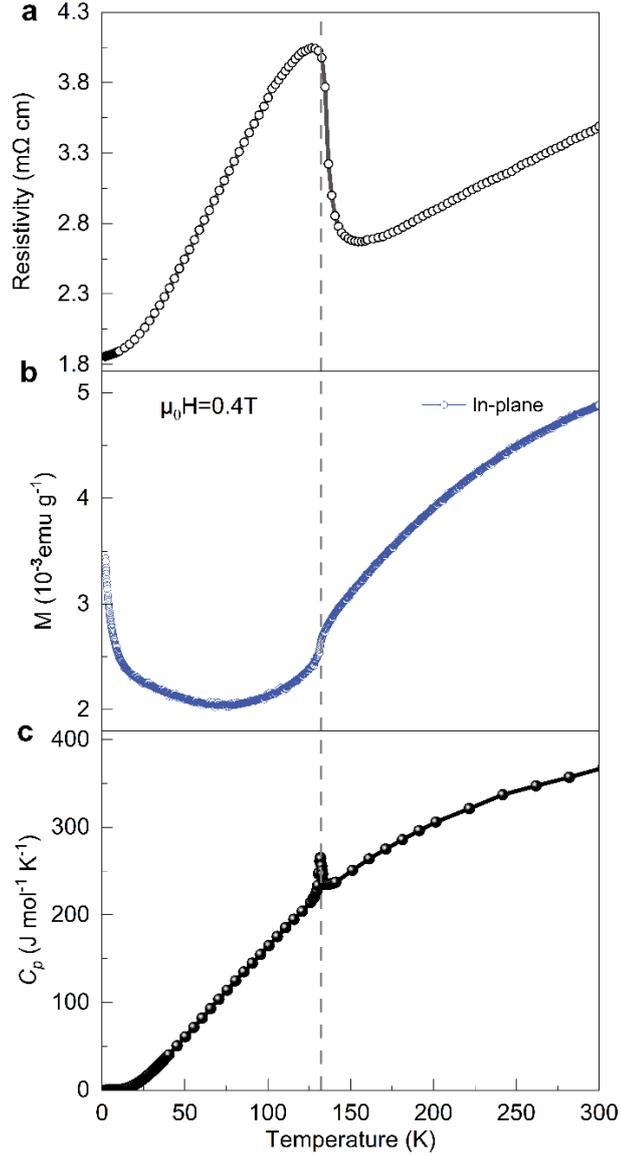

**FIG. 2. Physical properties of La$_4$Ni$_3$O$_{10}$ single crystals. a**, temperature-dependent resistivity. **b**, temperature-dependent magnetization. **c**, temperature-dependent heat capacity.

Furthermore, we investigate the electronic structure of La$_4$Ni$_3$O$_{10}$ using high-resolution angle-resolved photoemission spectroscopy (ARPES) (Fig. S4). Overall, the experimental band structure of La$_4$Ni$_3$O$_{10}$ is very similar to that of La$_3$Ni$_2$O$_7$ [30], despite their different crystal structure, suggesting unified electronic properties of nickelates in Ruddlesden-Popper phases.

## C. Superconductivity in pressurized single crystals.

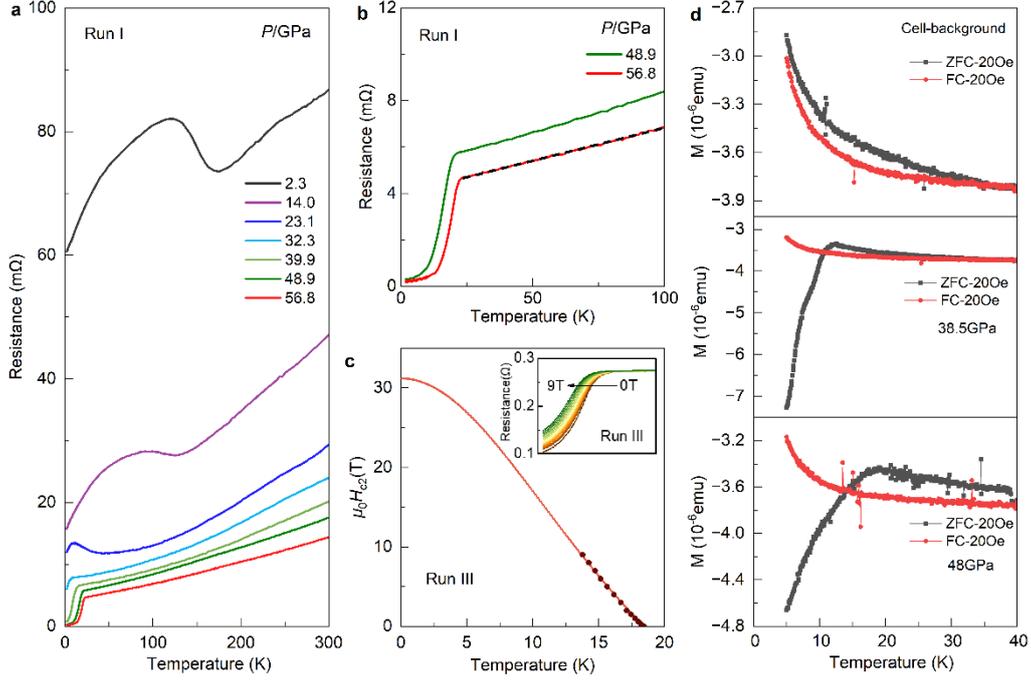

**FIG.3. Transport properties of $La_4Ni_3O_{10}$ single crystals as a function of pressure.** **a**, electrical resistance of $La_4Ni_3O_{10}$ as a function of temperature from 2.3 GPa to 56.8 GPa in Run I. **b**, linear resistance in the normal state. **c**, temperature dependence of the upper critical field $\mu_0H_{c2}(T)$. $T_c$ is determined as the 90% of the resistance at the onset $T_c$. The solid lines represent the Ginzburg-Landau (G-L) fitting. The inset shows the low-temperature resistance under various magnetic fields and the temperature dependence of the upper critical field for $La_4Ni_3O_{10}$ at 67.7 GPa in Run III. **d**, DC magnetic susceptibility of $La_4Ni_3O_{10}$ under high pressures.

The resistivity anomaly indicates that $La_4Ni_3O_{10}$ is located in the vicinity of an electronic instability. It is well known that superconductivity often appears near the critical point of structural or density waves. In this respect, pressure has been proven to be a convenient and effective way to modify lattice parameters and systematically influence the corresponding electronic states without introducing other factors. Motivated by these advantages, we measured the electrical resistance of $La_4Ni_3O_{10}$ single crystals at various pressures. Figure 3a shows the typical resistance curves of La$_4$Ni$_3$O$_{10}$ for pressure up to 60 GPa. The resistance anomaly is sensitive to external pressures. With increasing pressure to 2.3 GPa, the temperature of the resistance anomaly shifts to ~ 120 K. As the pressure continues to increase to 14.0 GPa, the anomalous feature in the resistance curve becomes broadened and less pronounced. At 23.1 GPa, the resistance anomaly is suppressed further and the resistance curve displays a slight upturn at the lower temperatures followed by a down-turn feature. When the pressure approaches 32.3 GPa, the temperature dependence of resistance changes to that of a normal metal. Interestingly, a distinct behavior characterized by a resistance drop of about 8.5 K emerges and this behavior becomes more pronounced as the pressure is increased, indicating the superconducting transition. The superconducting $T_c$ continuously increases and reaches 23.5 K at 56.8 GPa. Especially, with the appearance of superconductivity, *T*-linear resistance becomes clear gradually in the normal state (Fig. 3b). It is worth noting that the resistance drops by 95% from the normal state to the lowest temperature, which suggests a superconducting transition. Several independent high-pressure transport measurements on La$_4$Ni$_3$O$_{10}$ single crystals provide consistent and reproducible results (Figs. S5-6), confirming intrinsic superconductivity under high pressure. Compared with Run I, the resistance anomaly peak was suppressed below 10 GPa in Run II (Fig. S4), followed by the emergency of sharp drops. The superconducting $T_c$ reaches 25 K at approximately 42.5 GPa.

To further determine whether the observed resistance drop is truly associated with a superconducting transition, we performed resistance measurements under various magnetic fields. As displayed in Fig.3c and Fig.S6, the resistance drop of La$_4$Ni$_3$O$_{10}$ is progressively suppressed to lower temperatures, which implies a superconducting transition. The onset of superconductivity seems insensitive to the external magnetic fields and the transition width becomes broad with increasing magnetic field, which is similar to high-$T_c$ cuprates. As shown in the inset of Fig. 3c, superconductivity survives even under magnetic field $\mu_0H$ = 9 T. The upper critical field $\mu_0H_{c2}$ can be well-fitted using the empirical Ginzburg-Landau formula $\mu_0H_{c2}(T) =$ $\mu_0 H_{c2}(T)(1-t^2)/(1+t^2)$, where $t = T/T_c$. The superconducting transition is robust and extrapolated upper critical field $\mu_0 H_{c2}(0)$ of $La_4Ni_3O_{10}$ can reach 31.1 T at 67.7 GPa in Run III, which yields a Ginzburg-Landau coherence length $\xi_{GL}(0)$ of 3.25 nm. To further validate pressure-induced superconductivity, we utilized a specially designed micro-fabricated beryllium-copper alloy Diamond Anvil Cell (DAC) to conduct ultra-sensitive DC magnetic susceptibility measurements under high pressure (Fig. 3d and Fig. S7), with the aim of observing the diamagnetic effect characteristic of superconductors. From the raw data of high-pressure DC magnetic susceptibility, no significant transitions were detected in the background data of the DAC, regardless of whether the measurements were performed in Zero-Field-Cooling (ZFC) or Field-Cooling (FC) modes. However, when neon was employed as the pressure-transmitting medium and the sample pressure was increased to 38.5 GPa, a distinct diamagnetic transition was observed below 12.3 K in ZFC mode, confirming bulk superconductivity with a volume fraction exceeding 80%. The superconducting transition temperature derived from these magnetic measurements at this pressure was in agreement with the results obtained from electrical transport measurements, further corroborating the emergence of superconductivity under high pressure. When the pressure was increased to 48.0 GPa, the superconducting transition temperature was enhanced, with diamagnetic below 19 K, consistent with the superconducting transition phase diagram of the sample under pressure. In summary, through high-pressure DC magnetic susceptibility measurements, we directly observed the diamagnetic effect of superconductors under high pressure, thereby substantiating that the sample exhibits superconductivity under such conditions.

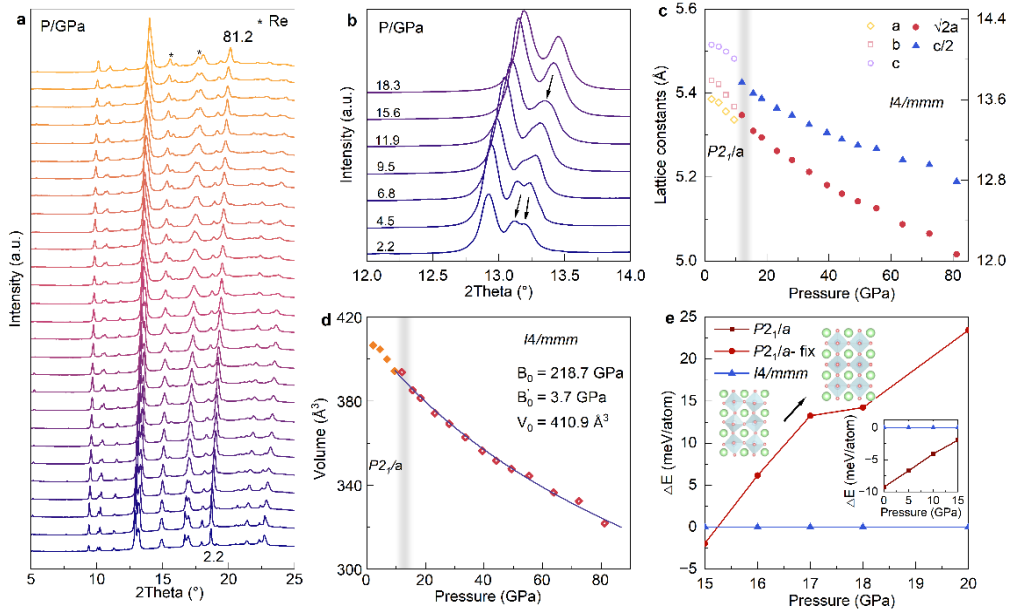

**FIG .4. Structural transition of $La_4Ni_3O_{10}$ single crystals as a function of pressure. a**, synchrotron XRD patterns of powder samples from 2.2 GPa to 81.2 GPa. The asterisk refers to the diffraction peaks of the Re gasket. **b**, details of the evolutions of the (0 2 0) and (-2 0 1) peaks under pressure. **c**, the evolution of lattice parameters refined from the XRD patterns. **d**, the pressure dependence of volume of $P2_1/a$ and $I4/mmm$ space group derived from the XRD refinement results. **e**, the circles represent the difference in enthalpy between the space groups $P2_1/a$ and $I4/mmm$ as a function of pressure calculated using the first-principles method. The inset shows the modification of $NiO_6$ octahedra stacking mode under pressure.

To clarify the emergency of superconductivity, we perform *in situ* high-pressure XRD measurements up to 80 GPa. As shown in Fig. 4a, except for two peaks arising from the Re gasket, all peaks could be indexed well to $La_4Ni_3O_{10}$ with the monoclinic structure. With the pressure increasing, both the lattice parameters and unit cell volume decrease continuously (Fig. 4c and 4d). At around 12 GPa, the (0 2 0) and (-2 0 1) peaks merge into one peak, suggesting the occurrence of structural phase transition (Fig. 4b and Fig. S8). The structural analysis revealed that *in situ* high pressure XRD patterns above 12 GPa can be well described with the tetragonal $I4/mmm$ space group (No. 139).

The enthalpy difference relative to $I4/mmm$ phase within 20 GPa is plotted in Fig. 4e. From 0 GPa to 15 GPa [inset of Fig. 4e], we optimized the ions' positions, cell shape, and cell volume, and the enthalpy difference between the $P2_1/a$ phase and $I4/mmm$ phase decreases with pressure. When pressure is higher than 15 GPa, the $P2_1/a$ phase transforms to $I4/mmm$ phase after structural optimization owing to the sensitivity of the $NiO_6$ octahedra stacking angles. Hence, we compress the volume with the ions' positions and volume shape fixed based on the optimized $P2_1/a$ structure at 15 GPa for the enthalpy calculations above 15 GPa. The enthalpy of $P2_1/a$ phase is higher than that of $I4/mmm$ phase after 15.2 GPa, illustrating that $I4/mmm$ phase is more energetically stable than $P2_1/a$ phase under high pressure, which is in agreement with *in situ* high pressure XRD measurements. Accompany by the phase transition, the Ni-O-Ni angles change within a layer, which leads to the buckled Ni-O planes becoming more flattened (Inset of Fig. 4e). Similar to cuprates, the appearance of a quasi-two-dimensional structure under pressure paves the way for the emergence of high-temperature superconductivity in nickelates.

Based on the above high-pressure characterizations, we construct the temperature-pressure phase diagram of $La_4Ni_3O_{10}$ as shown in Fig. 5a. In lower pressure region, the $La_4Ni_3O_{10}$ single crystal exhibits metal-to-metal transition with a density-wave resistance anomaly. As the pressure increases, the resistance anomaly is gradually suppressed. Upon further increasing pressure, superconductivity appears between 10-20 GPa, and the superconducting transition temperature $T_c$ reaches 25 K over 50 GPa. In addition, concomitant with the enhancement of superconductivity in this pressure range, the strange metal behavior becomes gradually evident above 30 GPa. As can be seen from the phase diagram, the competition between resistance anomaly and superconductivity resembles the situation in some change or spin density wave materials. All these observations indicate that high-temperature superconductivity can also be achieved in the pressurized $La_4Ni_3O_{10}$ single crystals.

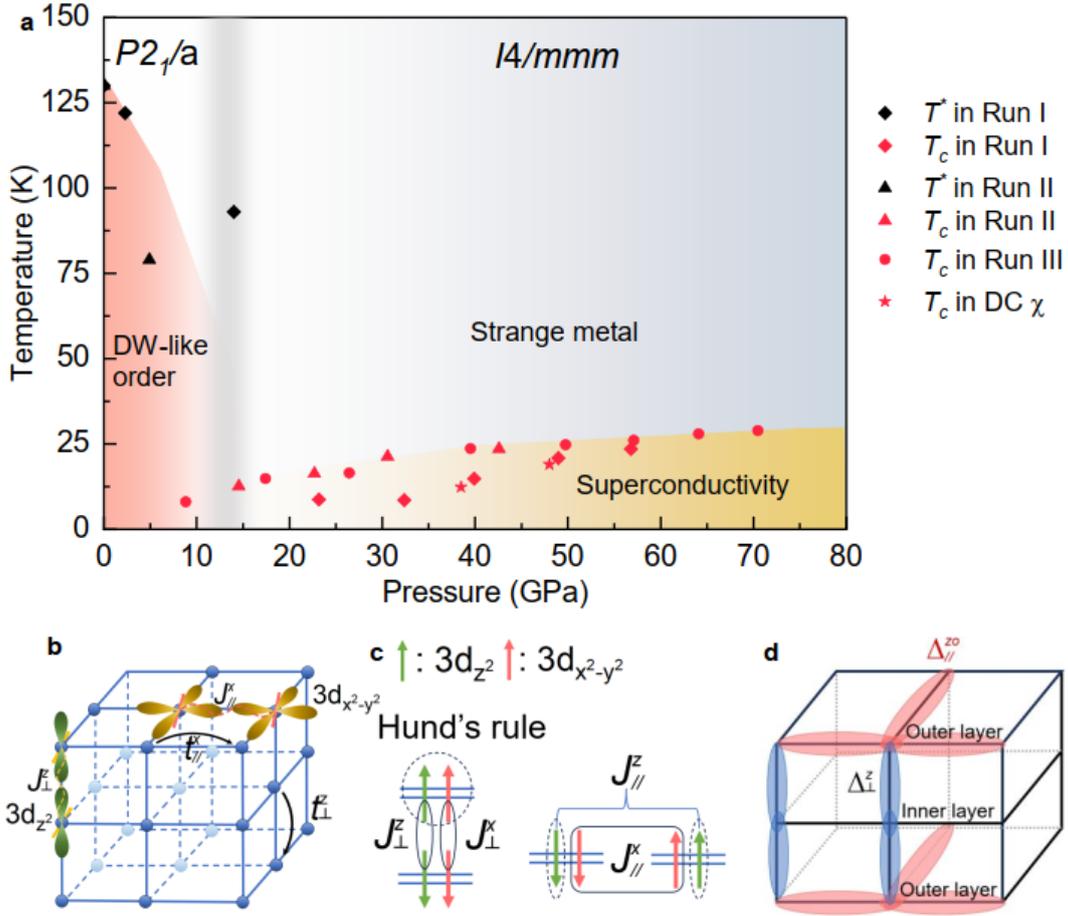

**FIG .5. The phase diagram and theoretical model of pressurized La$_4$Ni$_3$O$_{10}$. a**, phase diagram under pressure. The yellow symbols represent the density-wave (DW) transition $T^*$ measured at various pressures. The red symbols represent the superconducting transition temperatures $T_c$ determined from the various runs. **b**, the relevant orbitals, their couplings, and direct super-exchange interactions $J_\perp^z$ and $J_{//}^x$. **c**, an illustration of the effective super-exchange interactions transferred from one orbital to the other induced by Hund's rule coupling. Note that such transfer only takes place when both orbitals are singly occupied, while in the realistic Ni-3d$^{7.33}$ configuration, these orbitals have only a 2/3 possibility of being singly occupied. **d**, the dominant superconducting pairings.

### D. Theoretical analysis.

The pairing mechanism in pressurized La$_4$Ni$_3$O$_{10}$ is analyzed as follows. As shown in Fig. 5b, each unit cell contains three Ni- atoms located on three different layers, with each Ni- atom hosting two 3d orbitals, i.e. the $3d_{z^2}$ and $3d_{x^2-y^2}$. For this system, a six-orbital tight-binding (TB) model has been established, with the TB hopping integrals given in Ref. [21][21], [47][47], [48][48], [49][49], [50][50]. Among these hopping integrals, the strongest one is the interlayer hopping between the $3d_{z^2}$ orbitals, i.e. $t_\perp^z$, and the next strongest one is the intralayer hopping between the $3d_{x^2-y^2}$ orbitals, i.e. $t_{//}^x$. Under the strong Hubbard repulsion, these hopping integrals can induce the strongest AFM interlayer super-exchange $J_\perp^z$; and the next strongest AFM intralayer super-exchange $J_{//}^x$. In addition, the Hund's rule aligns spins in two orbitals on the same site. Hence, it transfers the nearest-neighbor super-exchange interaction from one orbital to the other [36], as shown in Fig. 5c, but with considerably reduced averaged strength because the transfer only takes place when the two orbitals on the same site are both singly occupied, while in the realistic Ni-$3d^{7.33}$ configuration, they have only 2/3 possibility to be singly-occupied[21]. Similar to the cuprates, these super-exchange interactions can serve as effective pairing interactions in pressurized $La_4Ni_3O_{10}$.

Based on the above analysis, a two-orbital t-J model has been established by some of the authors here in Ref. [51][51]. Then systematic slave-boson-mean-field approach is engaged to study this model. The obtained results suggest that while both orbitals are near 1/3-filling [21], the $3d_{z^2}$ orbital dominates the pairing because its weak intralayer hopping leads to weak band dispersion and hence enhances the density of state, which enhances the pairing strength. It is also shown in Ref. [51][51] that the lower $T_c$ in pressurized $La_4Ni_3O_{10}$ than that in pressurized $La_3Ni_2O_7$ originates from the pairing frustration effect caused by its trilayer structure. In bilayer $La_3Ni_2O_7$, driven by the interlayer super-exchange interaction, the pairing is mainly between the two layers [33, 34, 36][33, 34, 36], experiencing no frustration. However, in trilayer $La_4Ni_3O_{10}$, as shown in Fig. 5d, when an inner-layer $3d_{z^2}$ electron pairs with an upper outer-layer $3d_{z^2}$ electron driven by $J_\perp^z$, the lower outer-layer $3d_{z^2}$ electron is left single. That electron has to seek for pairing with neighboring electrons within the same layer. Since the intralayer super-exchange interaction $J_{//}^{z}$ for 3d$_{z^2}$ electron is weak, such a pairing is fragile and easily destroyed by thermal fluctuation. This pairing frustration effect in pressurized La$_4$Ni$_3$O$_{10}$ leads to a much lower $T_c$ than that in pressurized La$_3$Ni$_2$O$_7$, as shown in Fig.3 a and Fig.5a in Ref. [51][51]

## III. CONCLUSION

We have successfully grown high-quality single crystals of La$_4$Ni$_3$O$_{10}$ and performed a series of transport measurements both at ambient pressure and high pressure. We found pressure-induced superconductivity with a $T_c$ at approximately 25 K. The emergency of superconductivity accompanied by the phase transition from $P2_1/a$ to $I4/mmm$, leads to the buckled Ni-O planes become flatten ultimately. Theoretical analysis displays the AFM super-exchange interactions can serve as the effective pairing interaction for the emergence of superconductivity in pressurized La$_4$Ni$_3$O$_{10}$ single crystals. Our results demonstrate that high-temperature superconductivity can be achieved in other members of RP phases, from which we can explore the unconventional superconductivity in nickelates.

*Note added*: After we submitted to preprint (arXiv: 2311.07423), we became aware that there was another paper by Zhu et al. (arXiv:2311.07353) posted on arXiv on the same day[52]. They reported resistance measurements on La$_4$Ni$_3$O$_{10}$ single crystals, which is consistent with our data.

## ACKNOWLEDGEMENTS

We are grateful for the stimulating discussions with Chen Lu and Zhiming Pan. This work was supported by the National Natural Science Foundation of China (Grant No. 52272265), the National Key R&D Program of China (Grant Nos. 2023YFA1607400 and 2018YFA0704300). D.P. and Q. Z. acknowledge the support from Shanghai Key

Laboratory of Material Frontiers Research in Extreme Environments, China (No. 22dz2260800), the Shanghai Science and Technology Committee, China (No. 22JC1410300). L.X.Y. acknowledges the financial support from the National Key R&D Program of China (Grants No. 2022YFA1403200 and 2022YFA1403100) and the National Natural Science Foundation of China (Grant No. 12275148). S.C.Y. acknowledges the financial support from the National Key R&D program of China (2020YFA0309602, 2022YFA1402703). F.Y. acknowledges the financial support from the National Natural Science Foundation of China (Grant No. 12074031). H.J.G acknowledges the financial support from the National Natural Science Foundation of China (Grant No. 12004270) and Guangdong Basic and Applied Basic Research Foundation (Grant No. 2022B1515120020). C.J.W. is supported by the National Natural Science Foundation of China (Grants No. 12234016 and 12174317) and the New Cornerstone Science Foundation. The authors thank the Analytical Instrumentation Center (# SPST-AIC10112914), SPST, ShanghaiTech University. The authors thank the staff from BL15U1 at Shanghai Synchrotron Radiation Facility for assistance during data collection.

# APPENDIX: MATERIALS AND METHODS

## 1. Material synthesis and characterization

Single crystals of $La_4Ni_3O_{10}$ were grown by the high-pressure optical floating zone technique. Raw materials of $La_2O_3$ (99.99%) were dried at 900 °C overnight before the reaction, and then mixed with an appropriate amount of NiO. The mixture was ground thoroughly, pressed into pellets, and sintered at 1200 °C for 2 days with several intermediate grindings. The powders were pressed into seed and feed rods with a diameter of about 6 mm and sintered at 1300 °C for 2 hours. Subsequent single crystal growth was performed in a high-pressure optical floating zone furnace (HKZ-300) with a growth rate of 5 mm/h and 20 bar of oxygen. Single crystals larger than 5 mm in size can be obtained. The Powder X-ray diffraction (PXRD) patterns were taken using a Bruker D2 phaser with Cu-$K_\alpha$ radiation ($\lambda$ = 1.5418 Å) at room temperature. Energy-dispersive X-ray spectroscopy (EDS) was employed to determine the compositions of single crystals. The high-resolution scanning transmission electron microscopy (TEM) imaging was taken in JEM-ARM300F GRAND ARM.

## 2. STM measurements

$La_4Ni_3O_{10}$ single crystals were cleaved at 77 K with a background pressure of $2\times10^{-10}$ Torr and then were immediately transferred into the STM head for measurements. STM experiments were performed within a Unisoku low-temperature STM. All of the STM data were obtained at 4.3 K.

## 3. ARPES measurements

High-resolution angle-resolved photoemission spectroscopy (ARPES) experiments were conducted at beamline 03U in Shanghai Synchrotron Radiation Facility (SSRF). The samples were cleaved in situ under ultra-high vacuum below $7\times 10^{-11}$ mbar. Data were collected with a Scienta DA30 electron analyzer. The total energy and angular resolutions were set to 20 meV and 0.2°, respectively.

## 4. Transport properties under ambient and high pressure

Resistivity, heat capacity, and magnetic susceptibility under ambient pressure were measured using the Physical Property Measurement System (Dynacool, Quantum Design) and SQUID vibrating sample magnetometer (MPMS3, Quantum Design). High-pressure electrical transport measurements were performed in a diamond anvil cell (DAC) with culets of 200 μm and 300 μm. The cubic BN/epoxy mixture was used as an insulator layer between the BeCu gasket and electrical leads. Four Pt foils were arranged according to the van der Pauw method[53, 54]. We use NaCl as a pressure-transmitting medium (PTM) to improve the hydrostatic environment. Ultrasensitive magnetic susceptibility measurements were performed utilizing a bespoke beryllium-copper alloy miniature diamond anvil cell (DAC), fitted with a non-magnetic rhenium gasket. The measurements were carried out using the Magnetic Property Measurement System (MPMS3, Quantum Design). The DAC comprises a pair of diamond anvils, each with a diameter of 300 μm, which confine the sample within a chamber of 240 μm in diameter. This chamber housed a single crystal sample of La$_4$Ni$_3$O$_{10}$, measuring approximately 210 μm in diameter and 33 μm in thickness. Neon gas was used as the pressure transmitting medium to ensure an optimal hydrostatic pressure environment. The superconducting volume fraction calculation method is derived from ref[55][55](see the Supporting Information for details). *In situ* high-pressure XRD measurements were performed at the beamline 15U in the Shanghai Synchrotron Radiation Facility (λ=0.6199 Å) and Mineral oil was used as PTM. The pressure was determined by the ruby luminescence method[56].

## 5. First-principles calculations

First-principles band structure calculations were performed using Vienna ab initio simulation package (VASP)[57, 58] with a plane wave basis. Band structure: The exchange-correlation energy was considered under Perdew-Burke-Ernzerhof (PBE) type generalized gradient approximation (GGA)[59] with spin-orbit coupling included. Hubbard $U$ = 4.0 eV was applied to describe the localized 3d orbitals of Ni atoms. The cutoff energy for the plane-wave basis was set to 400 eV. A Γ-centered Monkhorst-Pack k-point mesh of 19×19×5 was adopted in the self-consistent calculations. The enthalpy difference calculation: The calculations use projector-augmented wave (PAW)[60] approach to describe the core electrons and their effects on valence orbitals and consider the $5s^2\ 5p^6\ 5d_1\ 6s_2$ for La, $3d^8\ 4s^2$ for Ni and $2s^2\ 2p^4$ for O. We set the plane-wave kinetic-energy cutoff to 700 eV, and the Brillouin zone is sampled with the special k-mesh generated by the Monkhorst-Pack scheme with a k-point spacing of 2π ×0.025 Å$^{-1}$. The convergence tolerance is 10$^{-6}$ eV for total energy and all forces are converged to be less than 0.003 eV/Å. At ambient pressure, the calculated volume is less than 1% larger than the experimental data, suggesting the reliability of the calculation.